\documentclass[preprint,12pt,sort&compress]{elsarticle}
\usepackage{amsfonts}
\usepackage{amsmath}
\usepackage{amsthm}
\usepackage{amssymb}
\usepackage{graphicx}
\usepackage{dcolumn}
\usepackage{bm}
\usepackage{bbding}
\usepackage{mathrsfs}
\usepackage{graphicx}

\biboptions{numbers,sort&compress}

\begin{document}

\begin{frontmatter}
\title{Quantum correlation cost of the weak measurement}
\author{Jun Zhang}
\author{Shao-xiong Wu}
\author{Chang-shui Yu\corref{cor1}}
\ead{quaninformation@sina.com}
\cortext[cor1]{Corresponding author. Tel: +86 41184706201}
\address{School of Physics and Optoelectronic Technology, Dalian University of
Technology, Dalian 116024, China}
\begin{abstract}
Quantum correlation cost (QCC) characterizing how much quantum correlation is used in a weak-measurement process is presented  based on the
trace norm.  It is shown that the QCC is related to the
trace-norm-based quantum discord (TQD) by only a factor that is determined
by the strength of the weak measurement, so it only catches partial quantumness of a quantum system compared
with the TQD. We also find that the residual quantumness
can be `extracted' not only by the further von Neumann measurement, but also
by a sequence of infinitesimal weak measurements. As an example, we
demonstrate our outcomes by the Bell-diagonal state.
\end{abstract}

\end{frontmatter}

\section{Introduction}

As one of the important quantum correlations, quantum entanglement has been
identified as an important physical resource in quantum information
processing tasks (QIPTs) \cite{rmp}. But in some QIPTs such as the robust
quantum algorithm against the decoherence \cite{robust}, the deterministic
quantum computation with one quantum bit (DQC1) \cite{DQC1} etc, is there
not any quantum entanglement, but quantum discord that was introduced in
Ref. \cite{Olliver} and Ref. \cite{Vedral}, respectively, has been shown to
be able to grasp more quantumness than entanglement. In recent years, the
research on quantum discord has been made great progress in various fields
\cite%
{zong,M,S,A,Y,yb,shang,shang1,yanhua,yanhua1,yanhua2,jieshi,jieshi1,jieshi2,celiang,celiang2,wxg,v1}%
.

Quantum discord was originally defined by the difference between the total
correlation and the classical correlation that is obtained by the optimal
local measurements, since classical information is locally accessible, and
can be obtained without perturbing the state of the system \cite{Olliver}.
Later it was generalized to various cases by using different distance
measures between the state taken into account and the post-measured state
\cite{entry, jihe, trace, dtd, baxi, luo,luo1, pro, hu, ado, Xiudis, Lina,
prl,jia,celiang1}. In usual, these mentioned measurements are referred as to
the von Neumann measurement which can cause collapse of the wave function
and hence has \textit{strong} influences on (destroys) the initial state. In
1988, Aharonov \textit{et al.} introduced the weak measurement \cite{weak}
which was applied in many areas \cite{bo, bo1, sin, cwe}. It was later
generalized by Oreshkov and Brun \cite{weak1} to the case with only
preselection in terms of measurement operator formalism and one of the
important characteristics is that the strength of the measurement process
can be controlled to be very weak, so the quantum state can be influenced
weakly. Considering such weak measurements in the quantification of quantum
correlation, Singh and Pati proposed the super quantum discord \cite{super}.
It has been shown that super discord is not less than the quantum discord.
Recently, the super quantum discord has attracted increasing interests \cite%
{ling, dandiao, fei,fanheng1,super1}. In particular, it is surprising that
the weak measurement can extract extra quantumness and the lost quantumness
could even be resurrected by weak measurement \cite{super1}.

In this paper, we find very different phenomena from those in Refs. \cite%
{super,super1}. Motivated by the super quantum discord, we present the quantum correlation cost (QCC) by
employing the trace norm as the distance measure between the state of
interests and post-weak-measured state. It is interesting that the
difference between the QCC and the quantum discord
based on trace norm (TQD) is in a factor that is determined by the strength
of the employed weak measurement. Instead of the extra quantumness \cite%
{super}, this factorization relation shows that the QCC quantifies how much
quantumness of a quantum system is `extracted' (used) by the employed weak
measurements. In particular, we also find that the residual quantumness
which the weak measurement fails to extract will be extracted further, if
the post-weak-measured state is succeeded by the von Neumann measurement or
by a sequence of infinitesimal weak measurements. In addition, all these
conclusions can also be generalized to $\left( 2\otimes d\right) -$
dimensional systems and the case of multiple-outcome weak measurements. As a
demonstration, we study the QCC of the Bell-diagonal state, which shows the
consistency with our conclusions.

\section{Quantum correlation cost and Residual quantumness}

\textit{Quantum correlation cost.-} To begin with, we would like to first
introduce the original super quantum discord $Q_{w}(\rho _{AB})$ which is
defined as%
\begin{equation}
Q_{w}(\rho _{AB})=\min_{\Pi _{i}^{A}}S_{w}(B|\{P^{A}(x)\})+S(\rho
_{A})-S(\rho _{AB}),
\end{equation}%
where $S(\cdot)$ represents the von Neumann entropy and
\begin{equation}
S_{w}(B|\{P^{A}(x)\})=p(+x)S(\rho _{B|P^{A}(+x)})+p(-x)S(\rho
_{B|P^{A}(-x)}),
\end{equation}%
with $\rho _{B|P^{A}(\pm x)}$ the post-weak-measured state given by%
\begin{equation}
\rho _{B|P^{A}(\pm x)}=\frac{Tr_{A}[(P^{A}(\pm x)\otimes I_{B})\rho
_{AB}(P^{A}(\pm x)\otimes I_{B})]}{Tr_{AB}[(P^{A}(\pm x)\otimes I_{B})\rho
_{AB}(P^{A}(\pm x)\otimes I_{B})]},  \label{3}
\end{equation}%
and $p(\pm x)$ the corresponding probability given by%
\begin{equation}
p(\pm x)=Tr_{AB}[(P^{A}(\pm x)\otimes I_{B})\rho _{AB}(P^{A}(\pm x)\otimes
I_{B})].
\end{equation}%
It is noted that%
\begin{equation}
\left\{
\begin{array}{c}
P(+x)=\alpha \pi _{1}+\beta \pi _{2} \\
P(-x)=\beta \pi _{1}+\alpha \pi _{2}%
\end{array}%
\right. ,
\end{equation}%
where $P^{\dagger }(+x)P(+x)+P^{\dagger }(-x)P(-x)=\mathbf{I}$ denotes the
two-outcome weak measurement with $\pi _{i}$ the normal projectors,
\begin{equation}
\alpha =\sqrt{\frac{1-\tanh x}{2}},\beta =\sqrt{\frac{1+\tanh x}{2}}.
\label{6}
\end{equation}%
and $x$$\in$
{}$\mathbb{R}${}
the strength of measurement process. It is obvious that the weak measurement
operators will be reduced to orthogonal projectors with $x\rightarrow \infty$.

The characteristic of the super quantum discord (SD) is that the SD can catch `extra' quantumness compared with the initial quantum discord. However, it is opposite to the original intention of quantum discord which requires to remove the classical correlation from the total correlation as much as possible by choosing the optimal measurements. It is obvious the weak measurement is the bad choice which cannot extract enough classical correlation. That is, the super quantum discord should include the residual classical correlation, so it seems to be greater than the usual quantum discord. However, the super quantum discord inspires us to study quantum correlation in a different way, that is, to be effectively related to the weak-measurement process. To do so, we define the quantum correlation cost (QCC) based on the trace norm as some kind of quantum correlation measure which is given as follows.

\textbf{Definition 1.} The QCC $D_{w}(\rho _{AB})$ for a bipartite quantum
state $\rho _{AB}$ is defined as%
\begin{equation}
D_{w}(\rho _{AB}):=\min_{\pi }\left\Vert \rho _{AB}-\Pi _{2}(\rho
_{AB})\right\Vert _{1},  \label{7}
\end{equation}%
where $\left\Vert X\right\Vert _{1}=Tr\sqrt{XX^{\dagger }}$ and $\Pi
_{2}(\cdot )$ denotes the operator of two-outcome weak measurement on
subsystem $A$ with
\begin{equation}
\Pi _{2}(\rho _{AB})=P(+x)\rho _{AB}P^{\dagger }(+x)+P(-x)\rho
_{AB}P^{\dagger }(-x).  \label{ad}
\end{equation}

It is implied that the QCC is defined in terms of the two-outcome weak
measurement, which holds throughout this paper if no particular statements
given. In addition, one can easily find that $D_{w}(\rho _{AB})$ inherits
the advantage of the trace norm such as the invariance under local unitary
operations and the contractivity under the local non-unitary evolution on
subsystem $A$, thus it is a reliable measure of quantum correlation. In
particular, one should note that the TQD $D(\rho _{AB})$ for $\rho _{AB}$
can be directly obtained by requiring $x\rightarrow \infty $, that is,
\begin{equation}
D(\rho _{AB})=\lim_{x\rightarrow \infty }D_{w}(\rho _{AB}).  \label{8}
\end{equation}%
With Eq. (\ref{7}) and Eq. (\ref{8}), we can fortunately find the deeper
relation between the QCC and the TQD which is given by the following
rigorous way.

\textbf{Theorem 1.} For a bipartite quantum system of qubits $\rho _{AB}$
\begin{equation}
D_{w}(\rho _{AB})=(1-2\alpha \beta )D(\rho _{AB}).
\end{equation}%
with $\alpha $, $\beta $ defined by Eq. (\ref{6}) and only determined by the
strength of the measurement process.

\textbf{Proof.} Let's consider an arbitrary two-qubit state $\rho _{AB}$.
Suppose that the final state after the weak measurement is $\Pi _{2}(\rho
_{AB})$. Substitute Eq. (\ref{ad}) into Eq. (\ref{7}), we have%
\begin{eqnarray}
D_{w}(\rho _{AB}) &=&\min_{\pi }\left\Vert \rho _{AB}-\Pi _{2}(\rho
_{AB})\right\Vert _{1}  \notag \\
&=&\min_{\pi }\|\rho _{AB}-\left( \pi _{1}\rho _{AB}\pi _{1}+\pi _{2}\rho
_{AB}\pi _{2}\right)  \notag \\
&&-2\alpha \beta \left( \pi _{1}\rho _{AB}\pi _{2}+\pi _{2}\rho _{AB}\pi
_{1}\right) \|_{1}  \notag \\
&=&(1-2\alpha \beta )\min_{\pi }\left\Vert \pi _{1}\rho _{AB}\pi _{2}+\pi
_{2}\rho _{AB}\pi _{1}\right\Vert _{1},  \label{10}
\end{eqnarray}%
Insert Eq. (\ref{8}) into Eq. (\ref{10}), it follows that
\begin{equation}
D_{w}(\rho _{AB})=(1-2\alpha \beta )D(\rho _{AB}).
\end{equation}%
which completes the proof.\hfill $\blacksquare$

Theorem 1 shows us a very simple factorization relation between the QCC and
the TQD, which provides the important root for the next stories.

\textit{Residual quantumness.-} Based on the factorization relation given
above, one will obviously see that $D_{w}(\rho _{AB})\leq D(\rho _{AB})$ due
to the reduction factor $(1-2\alpha \beta )\leq 1$ with `$=$' satisfied for $%
x\rightarrow \infty $. Since the weak measurement influences the system more
weakly than the normal projective measurement, the distance from the state
of interests to the post-weak-measured state is naturally less than that
from the state to the post-projective-measured state. Therefore, compared
with the TQD $D(\rho _{AB}),$ it is shown that the QCC $D_{w}(\rho _{AB})$
can only grasp the partial quantumness instead of the extra quantumness.
Thus the residual quantumness can be written as%
\begin{equation}
\Delta =D(\rho _{AB})-D_{w}(\rho _{AB})=2\alpha \beta D(\rho _{AB}).
\end{equation}%
Below we will show that the residual quantumness can be further extracted if
we continue performing a projective measurement on the post-weak-measured
state.

\textbf{Theorem 2. }Let $\tilde{\rho}_{AB}$ denote the final state of $\rho
_{AB}$ after the optimal weak measurement such that $D_{w}(\rho
_{AB})=\left\Vert\rho _{AB}-\tilde{\rho}_{AB}\right\Vert_{1}$, then we have
\begin{equation}
D(\tilde{\rho}_{AB})=D(\rho _{AB})-D_{w}(\rho _{AB}).  \label{a}
\end{equation}

\textbf{Proof.} At first, one should keep in mind that $\tilde{\rho}%
_{AB}=\Pi _{2}(\rho _{AB})$ given by Eq. (\ref{ad}) can be further written
as that implied in Eq. (\ref{10}) based on the Proof of Theorem 1. In
addition, in order to distinguish the projectors from those in the weak
measurement, we denote the projectors operated on subsystem $A$ in the
normal projective measurement by be $\pi _{1}^{\prime } $ and $\pi
_{2}^{\prime }$. Therefore, in terms of the definition of TQD, one can
obtain $D(\tilde{\rho}_{AB})$ as%
\begin{equation}
D(\tilde{\rho}_{AB})=\min_{\pi ^{\prime }}\left\Vert \tilde{\rho}%
_{AB}-\sum_{i=1}^{2}\pi _{i}^{\prime }\tilde{\rho}_{AB}\pi _{i}^{\prime
}\right\Vert _{1}.  \label{14}
\end{equation}%
Since for $x\rightarrow \infty $, the weak measurement operated on $\tilde{%
\rho}_{AB}$ should become the normal projective measurement, one will arrive
at
\begin{eqnarray}
\lim_{x\rightarrow \infty }D(\tilde{\rho}_{AB}) &=&\min_{\pi ^{\prime
}}\left\Vert \sum_{j=1}^{2}\pi _{j}\rho _{AB}\pi _{j}-\sum_{i,j=1}^{2}\pi
_{i}^{\prime }\pi _{j}\rho _{AB}\pi _{j}\pi _{i}^{\prime }\right\Vert _{1}
\notag \\
&=&0.  \label{15}
\end{eqnarray}%
where $\pi _{1}$ and $\pi _{2}$ are the projectors in the weak measurement.
It is obvious that the equality in Eq. (\ref{15}) holds iff the set of the
projectors \{$\pi _{i}^{\prime }\}=\{\pi _{j}\}$. Thus the extremum
operation in Eq. (\ref{15}) can be omitted, and then Eq. (\ref{15}) can be
rewritten as%
\begin{equation}
D(\tilde{\rho}_{AB})=\left\Vert \tilde{\rho}_{AB}-\sum_{i=1}^{2}\pi _{i}%
\tilde{\rho}_{AB}\pi _{i}\right\Vert _{1}.  \label{16}
\end{equation}%
Substitute the expression of $\tilde{\rho}_{AB}$ into Eq. (\ref{16}), we
will arrive at%
\begin{eqnarray}
D(\tilde{\rho}_{AB}) &=&\left\Vert \pi _{2}\tilde{\rho}_{AB}\pi _{1}+\pi _{1}%
\tilde{\rho}_{AB}\pi _{2}\right\Vert _{1}  \notag \\
&=&2\alpha \beta \left\Vert \pi _{1}\rho _{AB}\pi _{2}+\pi _{2}\rho _{AB}\pi
_{1}\right\Vert _{1}.  \label{17}
\end{eqnarray}%
where we use $\pi _{i}\pi _{j}=\delta _{ij}\pi _{i}$. The weak measurement
in $\tilde{\rho}_{AB}$ is required to be optimal in the sense of $D_{w}(\rho
_{AB})=\left\Vert \rho _{AB}-\tilde{\rho}_{AB}\right\Vert _{1}$, so the
projectors in the weak measurement will also be optimal for the
corresponding TQD for $x\rightarrow \infty $, because if it is not the case,
it will lead to two different TQDs for the same $\rho _{AB}$, which is also
interpreted in Ref. \cite{super1}. That is,
\begin{eqnarray}
D(\rho _{AB}) &=&\left\Vert \rho _{AB}-\sum_{i=1}^{2}\pi _{i}\rho _{AB}\pi
_{i}\right\Vert _{1}  \notag \\
&=&\left\Vert \pi _{1}\rho _{AB}\pi _{2}+\pi _{2}\rho _{AB}\pi
_{1}\right\Vert _{1}.  \label{18}
\end{eqnarray}%
Eq. (\ref{17}) and Eq. (\ref{18}) show that
\begin{equation}
D(\tilde{\rho}_{AB})=2\alpha \beta D(\rho _{AB}).
\end{equation}%
which accompanied with Theorem 1 implies $D(\tilde{\rho}_{AB})=D(\rho
_{AB})-D_{w}(\rho _{AB})$. The proof is finished.\hfill$\blacksquare$

In Theorem 2, we have shown that the residual quantum correlation that the
weak measurement fails to `extract' can be `extracted' by the latter projective
measurement. In addition, it is implied that no extra quantumness is wasted
compared with one optimal projective measurement, even though we use weak
measurements and projective measurements, respectively. That is, the
summation of the QCC and the residual TQD is completely consistent with the
TQD. Since any projective measurement can be implemented by a sequence of
continuous weak measurements, one could naturally ask if the quantum
correlation can be extracted little by little by these weak measurements. In
the following theorem, we will show that the weak measurement can do this
job indeed.

\textbf{Corollary.} Suppose $\rho _{n}$ to be the final state after $n$
optimal weak measurements with the same infinitesimal measurement strength
on the subsystem $A$ of $\rho _{AB}$ such that $D_{w}(\rho _{n})=\left\Vert
\rho _{n}-\rho _{n+1}\right\Vert _{1}$ with $\rho _{0}=\rho _{AB}$, then we
will have%
\begin{equation}
D_{w}(\rho _{n})=(1-2\alpha \beta )\left( 2\alpha \beta \right) ^{n}D(\rho
_{AB}),
\end{equation}%
and
\begin{equation}
D(\rho _{AB})=\lim_{x\rightarrow 0}\sum_{n=0}^{\infty }D_{w}(\rho _{n}).
\end{equation}

\textbf{Proof.} According to Theorem 1, for $\rho _{n}$ we have
\begin{equation}
D_{w}(\rho _{n})=(1-2\alpha \beta )D(\rho _{n}).  \label{22}
\end{equation}%
Using Theorem 2 and its proof, one can find
\begin{equation}
D(\rho _{n})=D(\rho _{n-1})-D_{w}(\rho _{n-1})=2\alpha \beta D(\rho _{n-1}),
\end{equation}%
which directly leads to
\begin{equation}
D(\rho _{n})=\left( 2\alpha \beta \right) ^{n}D(\rho _{AB}).  \label{24}
\end{equation}%
Insert Eq. (\ref{24}) into Eq. (\ref{22}), we will obtain%
\begin{equation}
D_{w}(\rho _{n})=(1-2\alpha \beta )\left( 2\alpha \beta \right) ^{n}D(\rho
_{AB}).  \label{25}
\end{equation}%
Since we consider the infinitesimal measurement strength which means $%
x\rightarrow 0$, i.e., $\alpha \beta \rightarrow 0$, sum Eq. (\ref{25}) over
$n$, one will get%
\begin{eqnarray}
\lim_{x \rightarrow 0}\sum_{n=0}^{\infty }D_{w}(\rho _{n}) &=&\lim_{x
\rightarrow 0}(1-2\alpha \beta )D(\rho _{AB})\sum_{n=0}^{\infty }\left(
2\alpha \beta \right) ^{n}  \notag \\
&=&D(\rho _{AB}).  \label{26}
\end{eqnarray}%
Eq. (\ref{25}) and Eq. (\ref{26}) complete the proof.\hfill$\blacksquare$

Since only partial quantum correlation can be grasped by the weak
measurement, the above corollary first shows that the quantum correlation
can be `extracted' continuously by the infinitesimal weak measurements, which
is consistent with the fact that the continuous infinitesimal weak
measurements can realize the projective measurement. What is important is
that the series of weak measurements do not waste extra quantumness either.
On the contrary, if consider that the projective measurement will destroy
all the quantum correlation, in principle after the projective measurement,
the latter weak measurement will `extract' no quantum correlation. In the
following theorem, we will prove that it is the case.

\textbf{Theorem 3. }Let the final state of $\rho _{AB}$ after any projective
measurement on subsystem $A$ is given by $\tilde{\rho}_{AB}^{\prime
}=\sum_{j=1}^{2}\tilde{\pi}_{j}^{\prime }\rho _{AB}\tilde{\pi}_{j}^{\prime }$
, then we have
\begin{equation}
D_{w}(\tilde{\rho}_{AB}^{\prime })=0.
\end{equation}

\textbf{Proof.} Since $\tilde{\rho}_{AB}^{\prime }=\sum_{j=1}^{2}\tilde{\pi}%
_{j}^{\prime }\rho _{AB}\tilde{\pi}_{j}^{\prime }$, it is obvious that $D(%
\tilde{\rho}_{AB}^{\prime })=0$. According to Theorem 1, one can easily find
that $D_{w}(\tilde{\rho}_{AB}^{\prime })=(1-2\alpha \beta )D(\tilde{\rho}%
_{AB}^{\prime })=0$. The proof is completed.\hfill$\blacksquare$

\textit{The case of multiple-outcome weak measurement.-}One can find that
all the jobs presented above only cover the two-outcome weak measurement.
Next we will extend the weak measurement to the case of $n$ outcomes which
can be written as
\begin{equation}
P(i)=\alpha _{i}\pi _{1}+\beta _{i}\pi _{2},
\end{equation}%
where the real $\alpha _{i},\beta _{i}$ with $\sum_{i=1}^{n}\alpha
_{i}^{2}=1 $ and $\sum_{i=1}^{n}\beta _{i}^{2}=1$ are determined by the
strength vector $\vec{x}$ of measurement process and similarly $\pi _{1}$
and $\pi _{2}$ are the projectors satisfying $\pi _{1}+\pi _{2}=\mathbf{I}$.
In particular, one should note that in the limitation $\vec{x}\rightarrow
\infty $, $P(i)$ will become the normal projective measurement. Following
the same procedure as we've done for the two-outcome weak measurement, we
will have the following theorem.

\textbf{Theorem. 4}. All the above conclusions hold for $n$-outcome weak
measurement if the factor $2\alpha \beta $ corresponding to two outcomes is
replaced by $\sum_{i=1}^{n}\alpha _{i}\beta _{i}$ corresponding to the $n$
outcomes.

\textbf{Proof.} The proof is completely analogous to those given for the
two-outcome case, so it is omitted.\hfill$\blacksquare$

\textit{$(2\otimes d)$-dimensional systems.-} Finally, we would like to
emphasize that all the theorems as well as the corollary hold for $(2\otimes
d)$-dimensional systems, which can be confirmed if following the completely
analogous proof procedure for the two-outcome case.

To sum up, one can easily find that the QCC is quite different from the SD as well as the TQD. That is, the QCC
characterizes how much quantum correlation is 'extracted' (used) in the weak-measurement process. Even though the quantum correlation in a system can be 'extracted' by infinite weak measurements, the QCC subject to the weak measurement \textit{per se} does not describe how much correlation is present in the system.  As mentioned previously, the SD badly characterizes the classical correlation in a system, which can be seen when the measurement strength tends to zero ( the SD will reach its maximum, the total correlation and the QCC will reach its minimum, zero). The TQD describes the quantum correlation in the considered system.

\section{The application}

As a demonstration, let's consider the Bell-diagonal states given by
\begin{equation}
\rho _{AB}=\frac{1}{4}(\mathbf{I}_{AB}+\sum\limits_{k=1}^{3}c_{k}\sigma
_{k}^{A}\otimes \sigma _{k}^{B}),
\end{equation}%
with $\sigma _{k}$ the Pauli matrices and $\left\vert c_{3}\right\vert
>\left\vert c_{2}\right\vert >\left\vert c_{1}\right\vert $. Based on Refs.
\cite{trace,baxi,jia}, one can easily find that the QCC and TQD are given by
\begin{equation}
D(\rho _{AB})=\left\vert c_{2}\right\vert ,  \label{30}
\end{equation}%
and%
\begin{equation}
D_{w}(\rho _{AB})=(1-2\alpha \beta )\left\vert c_{2}\right\vert .  \label{31}
\end{equation}%
In particular, one can find that the optimal projectors that lead to the QCC
and TQD are $\hat{\pi}_{1}=\left\vert 0\right\rangle \left\langle
0\right\vert $ and $\hat{\pi}_{2}=\left\vert 1\right\rangle \left\langle
1\right\vert $. Eq. (\ref{30}) and Eq. (\ref{31}) show the consistency with
Theorem 1. Substitute the optimal projectors into the weak measurement, one
can obtain the weak measurement as $\hat{P}(x)=\alpha \hat{\pi}_{1}+\beta
\hat{\pi}_{2}$ and $\hat{P}(-x)=\beta \hat{\pi}_{1}+\alpha \hat{\pi}_{2}$.
So we can write the post-measured state as
\begin{eqnarray}
\tilde{\rho}_{AB} &=&\hat{P}(x)\rho _{AB}\hat{P}(x)+\hat{P}(-x)\rho _{AB}%
\hat{P}(-x)  \notag \\
&=&\frac{1}{4}(\mathbf{I}_{AB}+\sum\limits_{k=1}^{3}\tilde{c}_{k}\sigma
_{k}^{A}\otimes \sigma _{k}^{B}),
\end{eqnarray}%
with $\tilde{c}_{k}=2\alpha \beta c_{k}$ for $k=1,2$ and $\tilde{c}%
_{3}=c_{3} $. Use Eq. (\ref{17}) again, one can easily find that
\begin{equation}
D(\tilde{\rho}_{AB})=2\alpha \beta \left\vert c_{2}\right\vert ,
\end{equation}%
which associated with Eqs. (\ref{30},\ref{31}) demonstrates the validity of
Theorem 2. In addition, if we continue performing a weak measurement on $%
\tilde{\rho}_{AB}$, one will find that $\tilde{c}_{k}$ will be reduced
further by a factor $2\alpha \beta $ and $c_{3}$ will be still invariant. So
one will find $D(\tilde{\rho}_{2})=\left( 2\alpha \beta \right)
^{2}\left\vert c_{2}\right\vert $ and $D_{w}(\tilde{\rho}_{2})=(1-2\alpha
\beta )\left( 2\alpha \beta \right) ^{2}\left\vert c_{2}\right\vert $ with $%
\tilde{\rho}_{2}$ representing the twice weak measurements. This is
consistent with our corollary. Finally, let the projectors $\hat{\pi}_{1}$
and $\hat{\pi}_{2}$ be performed on the subsystem $A$ of $\rho _{AB}$, one
will get the final state as%
\begin{gather}
\tilde{\rho}_{AB}^{\prime }=\hat{\pi}_{1}\rho _{AB}\hat{\pi}_{1}+\hat{\pi}%
_{2}\rho _{AB}\hat{\pi}_{2}  \notag \\
=\frac{1}{4}\left[ \left\vert 0\right\rangle \left\langle 0\right\vert
\otimes \left( \mathbf{I}_{B}+c_{3}\sigma _{3}^{B}\right) +\left\vert
1\right\rangle \left\langle 1\right\vert \otimes \left( \mathbf{I}%
_{B}-c_{3}\sigma _{3}^{B}\right) \right] .
\end{gather}%
which obviously has no quantum correlation. The similar conclusion for $%
\tilde{\rho}_{AB}^{\prime }$ can also be found if one employs arbitrary
projectors instead of $\hat{\pi}_{1}$ and $\hat{\pi}_{2}$. Therefore,
Theorem 3 in this Bell-diagonal state is also satisfied.

\section{Conclusion and discussions}

We have presented the QCC for the weak measurement based on the trace norm. The QCC quantifies the
quantumness  `extracted'  by the employed weak measurements during the measurement procedure. We have found a
factorization relation between the QCC and the TQD. This is easily understood since
weak measurement only influences the system weakly. It is especially
interesting that the residual quantumness after the weak measurement can be
`extracted' further by the latter projective measurement or by the latter sequence of
infinitesimal weak measurements, which shows the consistent nature with that
the sequence of infinitesimal weak measurements can realize the projective
measurements. It is important that these different measurements do not waste extra quantumness. In contrast, it is shown that the weak measurement cannot
extract any quantum correlation from the state after projective
measurements. We believe that it provides a new point of view for us to understand
the weak measurement and the projective measurement. In addition, we also
generalize our conclusions to the cases of $2\otimes d$-dimensional systems
and of the multiple-outcome measurement. Finally, we demonstrate our
theorems and corollary by the Bell-diagonal states.

\section*{Acknowledgement}

This work was supported by the National Natural Science Foundation of China,
under Grants No.11375036 and 11175033.

\end{document}